\documentclass[msmath,amssymb,aps,pra,twocolumn,epsfig,showpacs,bibliography,
lengthcheck,superscriptaddress]{revtex4-1}

\usepackage{graphicx}
\usepackage{dcolumn}
\usepackage{bm}
\usepackage{hyperref}
\usepackage{epstopdf}
\usepackage{subcaption}
\usepackage{caption}
\begin{document}
\title{Interactions between non-resonant rf fields and atoms with strong spin-exchange collisions}
\author{C.-P. Hao}
\affiliation{Hefei National Laboratory of Physical Sciences at the Microscale, University of Science and Technology of China, Hefei 230026, China}
\affiliation{Department of Precision Machinery and Precision Instrumentation, University of Science and Technology of China, Hefei 230027, China}
\affiliation{Key Laboratory of Precision Scientific Instrumentation of Anhui Higher Education Institutes, University of Science and Technology of China, Hefei 230027, China}
\author{Z.-R. Qiu}
\author{Q. Sun}
\author{Y. Zhu}
\affiliation{Hefei National Laboratory of Physical Sciences at the Microscale, University of Science and Technology of China, Hefei 230026, China}
\author{D. Sheng}
\email{dsheng@ustc.edu.cn}
\affiliation{Hefei National Laboratory of Physical Sciences at the Microscale, University of Science and Technology of China, Hefei 230026, China}
\affiliation{Department of Precision Machinery and Precision Instrumentation, University of Science and Technology of China, Hefei 230027, China}
\affiliation{Key Laboratory of Precision Scientific Instrumentation of Anhui Higher Education Institutes, University of Science and Technology of China, Hefei 230027, China}

\begin{abstract}

We study the interactions between oscillating non-resonant rf fields and atoms with strong spin-exchange collisions in the presence of a weak dc magnetic field. We find that the atomic Larmor precession frequency shows a new functional form to the rf field parameters when the spin-exchange collision rate is tuned. In the weak rf field amplitude regime, a strong modification of atomic Larmor frequency appears when the spin-exchange rate is comparable to the rf field frequency. This new effect has been neglected before due to its narrow observation window. We compare the experimental results with density matrix calculations, and explain the data by an underdamped oscillator model. When the rf field amplitude is large, there is a minimum atomic gyromagnetic ratio point due to the rf photon dressing, and we find that strong spin-exchange interactions modify the position of such a point. 

\end{abstract}

\maketitle

\section{Introduction}
Interactions between non-resonant rf fields and atoms can modify the atomic Zeeman spectroscopy, which has been applied to resonant transfer of coherence~\cite{haroche70b} and precision measurements searching for new physics beyond the standard model~\cite{muskat87,esler07,sheng14a}. The pioneering work of Haroche {\it et al.}~\cite{haroche70a} showed that in a system with negligible collisions and rf fields oscillating perpendicular to the bias dc field, the atomic gyromagnetic ratio is modified through rf photon dressing. There are further investigations extended to cases with more complex orientations or components of the rf fields~\cite{yabuzaki72,yabuzaki74,summhammer93,bevilacqua12}.  Moreover, the Hamiltonian describing such interactions is  related to a class of dynamical control of the trap potentials in cold atom experiments~\cite{drese97,lignier07}. Eq.~\ref{eq:bessel} shows the modification of the gyromagnetic ratio \cite{haroche70a} through a functional dependence on rf field parameters when collisions are negligible.
\begin{equation}
\label{eq:bessel}
\gamma(B_{rf},\omega_{rf})=\gamma(0)J_0(\gamma(0)B_{rf}/\omega_{rf}),
\end{equation} 
where $\gamma(0)$ is atomic gyromagnetic ratio without external fields, $J_0$ is the zero order Bessel function, $B_{rf}$ and $\omega_{rf}$ are the amplitude and frequency of the rf field. 

However, spin-exchange interactions are normally the dominant collisions between atoms in alkali atom vapor cell experiments. This interaction conserves total angular momentum of the internal states. Recently it has been used for quantum manipulation to generate deterministic entanglement in a sample of ultracold atoms~\cite{Luo07}. When the spin-exchange collision rate $A_{se}$ is much larger than the atomic Larmor frequency, there exists a spin-exchange-relaxation-free (SERF) regime~\cite{happer73,allred02}. In this regime, the observed atomic gyromagnetic ratio is also modified due to the weighted average time of atoms staying in different hyperfine states~\cite{savukov05}, which in turn makes $\gamma(0)$ in Eq.~\ref{eq:bessel} smaller than the free atomic gyromagnetic ratio $\gamma_0$.
 
Connections develop between spin-exchange collisions and rf fields through their interactions with the atoms. Ref.~\cite{limes18} uses the resonant rf fields to effectively cancel  the phase evolution difference between ground hyperfine states, and extend the range of the SERF regime. In this paper, we perform atomic spectroscopy in the presence of non-resonant rf fields and spin-exchange interactions. We focus on the low bias field regime, where the atomic Larmor frequency is much smaller than the rf field frequency $\omega_{rf}$. By tuning the atomic density, we scan the spin-exchange collision rate over a large range. 

When the parameter $\gamma_0B_{rf}/\omega_{rf}$ is small, we could treat the rf fields as perturbations to the free evolution of the electron spins, and this is the weak rf field amplitude regime. We present the study of this regime in Sec. II, where a region of particular interest appears when $A_{se}$ is comparable to $\omega_{rf}$. There we find a surprisingly large modification of the atomic Larmor frequency. We also find the amplitude of this modification decreases rapidly as the bias field increases, which makes this effect only observable in a narrow window of experimental parameters. The experimental results are also  confirmed by density matrix simulations. We explain these results using a damped oscillator model, where the damping effect comes from the disturbance of the spin-temperature distribution determined by the strong spin-exchange collisions.

When $\gamma_0B_{rf}/\omega_{rf}>1$, atomic gyromagnetic ratio is strongly modified by the rf fields, and this is the large rf field amplitude regime. We present the study of this regime in Sec.~III, where we use a different analysis method compared with Sec.~II. There is a minimum atomic gyromagnetic ratio point due to the rf photon dressing in this regime. This point is determined by $J_0(\gamma_0B_{rf}/\omega_{rf})=0$ in the absence of spin-exchange interactions, and we find that strong spin-exchange interactions modify the position of such a point.

\section{Weak rf field amplitude regime}
\begin{figure}[htb]
\includegraphics[width=3in]{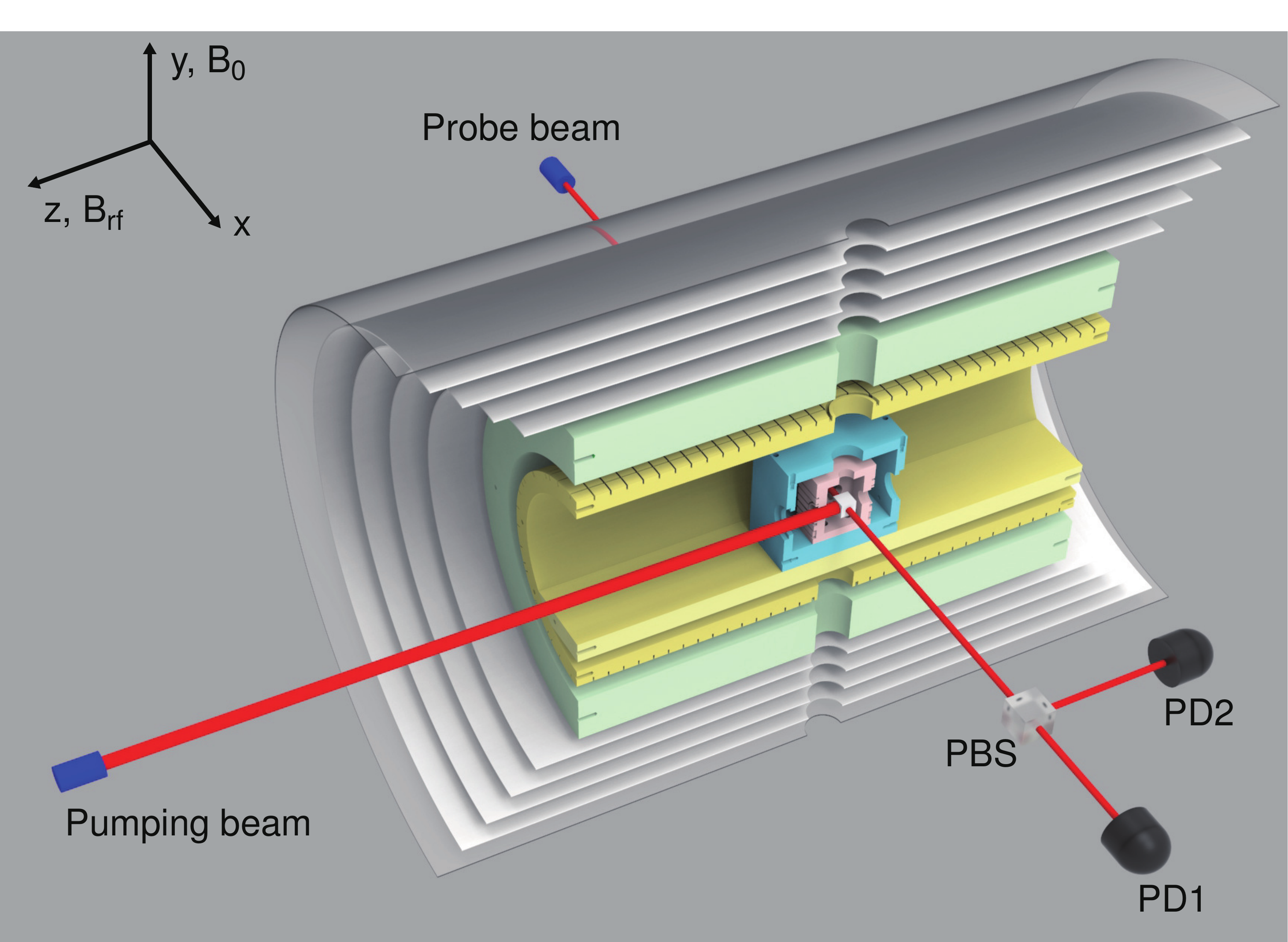}
\caption{\label{fig:setup}(Color online)  
A schematic plot of the experiment setup.} 
\end{figure}

We perform the experiments on cubic cells with an external size of 10~mm made from Pyrex glasses. The cells sit inside 5-layer cylindrical $\mu$-metal shields with the innermost layer length of 41~cm and diameter of 18~cm. An ac current heated oven is used to control the cell temperature. Fig.~\ref{fig:setup} shows the experimental setup. We have tested two atomic systems with different nuclear spin numbers. For $I=5/2$, we use a cell filled with $^{85}$Rb atoms and 600 torr N$_2$ gas. For $I=3/2$, we use a cell filled with $^{87}$Rb atoms, K atoms with natural abundance, 50 torr N$_2$ and 150 torr Ne gas. Each experiment cycle is 1 sec in length. In the first half cycle, we optically pump the alkali atoms at zero field using a circularly polarized beam on resonance with the corresponding atomic $D_1$ transition. In the second half cycle, we turn off the pumping beam, apply a bias field $B_0$ less than 1~mG in the $y$ direction, turn on a kHz rf field along the $z$ direction, and probe the spin precession signal with a far off-resonance linearly polarized light using the Faraday rotation effect.

\begin{figure}[htb]
\includegraphics[width=3in]{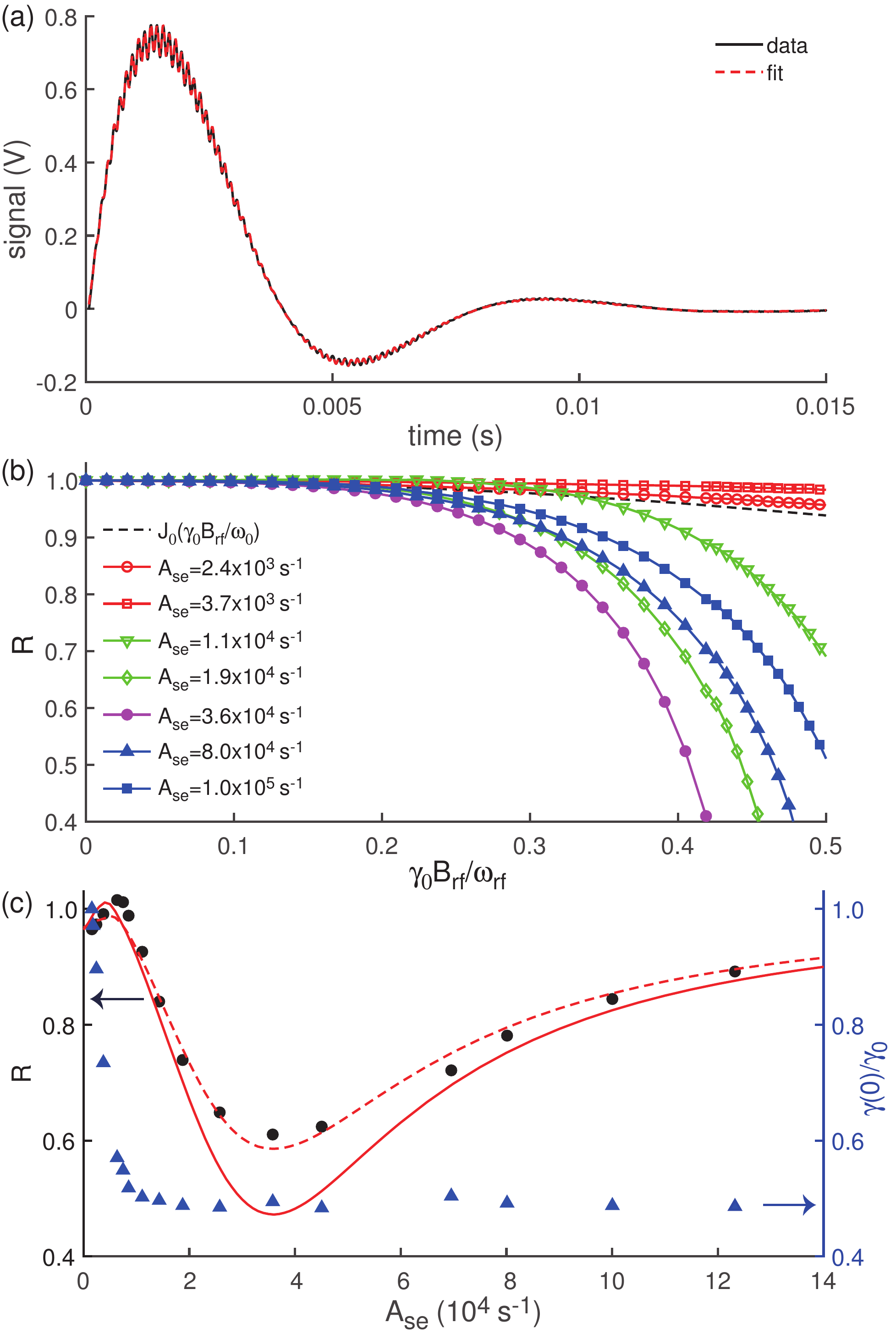}
\caption{\label{fig:Ase}(Color online) (a) A typical experimental data with a fitting curve.  The data is taken using the $I=3/2$ system with $B_0=0.3$~mG, $A_{se}=4.7\times10^4$~s$^{-1}$, $\omega_{rf}=2\pi\times4$~kHz, and $B_{rf}=2.6$~mG. (b) Experimental data of $R$ as a function of $\gamma_0B_{rf}/\omega_{rf}$ at different $A_{se}$, where $I=5/2$, $B_0=0.3$~mG, and $\omega_{rf}=2\pi\times4$~kHz.  (c) Experimental data (solid circles) and simulation results (lines) of $R$ as a function of $A_{se}$ at $\gamma_0B_{rf}/\omega_{rf}=0.391$ in plot~(a), where the dash line shows modified simulation result with a $B_{rf}$ 4\% smaller than the experimental parameter. This plot also includes the experimental data (triangle points) of $\gamma(0)/\gamma_0$ as a function of $A_{se}$.}
\end{figure}

Figure~\ref{fig:Ase}(a) shows a typical experimental data. In the presence of rf fields and spin-exchange interactions, the spin evolution is generally expressed by a Fourier series expansion. When $\gamma_0B_{rf}/\omega_{rf}$, introduced in Eq.~\ref{eq:bessel} as the modulation index parameter, is small, the effects of rf fields could be treated as perturbations, so that we can keep the leading terms and fit data using the equation:
\begin{eqnarray}\label{eq:fit}
y=&&ae^{-t/T}\sin(\omega_Lt+\phi_0)[1+c_1\sin(\omega_{rf} t+\phi_1)\nonumber\\
&&+c_2\sin(2\omega_{rf} t+\phi_2)]+b,
\end{eqnarray}
where $\omega_L$ is the Larmor precession frequency due to the bias field $B_0$, the terms with $\omega_{rf}$ and $2\omega_{rf}$ comes from modulation of the atomic polarization orientation due to the rf field. We define the parameter $R=\omega_L(\gamma_0B_{rf}/\omega_{rf})/\omega_L(0)$ ($\omega_L(0)=\gamma(0)B_0$) to characterize the interaction between atoms and rf fields, and probe the effects of spin-exchange collisions on this interaction from the changes in $R$ as we scan the spin-exchange collision rate $A_{se}$. In the experiment, we tune $A_{se}$ from $10^3$ to $10^5$ s$^{-1}$ by changing the cell temperature from 60 $^\circ$C to higher than 160 $^\circ$C. We calibrate $A_{se}$ by measuring $T_2$ at a large bias field in the low polarization limit, and using the nuclear slow-down factors~\cite{happer77}.

Figure~\ref{fig:Ase}(b) shows $R$ as a function of rf field parameters for atoms with nuclear spin $I=5/2$, $\omega_{rf}=2\pi\times$4~kHz, and $B_0=0.3$~mG. As we scan $A_{se}$, $R$ remains a monotonic function of $\gamma_0B_{rf}/\omega_{rf}$, but shows a complex relation with $A_{se}$, which deviates from the zero order Bessel function in Eq.~\ref{eq:bessel}. To better understand their relation, we fix $\gamma_0B_{rf}/\omega_{rf}=0.391$, and plot $R$ as a function of $A_{se}$ in Fig.~\ref{fig:Ase}(c). There are two turning points in this plot. When $A_{se}<6\times10^{3}$~s$^{-1}$, $R$ increases together with $A_{se}$. This is due to the fact that the system starts to enter the SERF regime, the gyromagnetic ratio without external rf fields $\gamma(0)$ decreases when $A_{se}$ increases as shown in the same plot, and $R$ in turn decreases slower as a function of $B_{rf}$ according to Eq.~\ref{eq:bessel}. For $A_{se}>6\times10^3$~s$^{-1}$, $\gamma(0)$ is stable but $R$ shows a non-monotonic relation with $A_{se}$, and a minimum value appears at $A_{se}\sim3.6\times10^4$~s$^{-1}$. This new spectroscopy result shows the strong influence of spin-exchange interactions on rf spectroscopy when $A_{se}$ and $\omega_{rf}$, the spin-exchange collision rate and the dressing rf field frequency, are comparable.

We simulate the spin dynamics in this experiment using the standard density matrix formalism~\cite{savukov05} 
\begin{eqnarray}\label{eq:denmat}
\frac{d\rho}{dt}&=&\frac{a_{hf}[\bm{I}\cdot\bm{S},\rho]}{i\hbar}+\frac{\mu_Bg_s[\bm{B}\cdot\bm{S},\rho]}{i\hbar}\nonumber\\
&+&A_{se}[\varphi(1+4\langle\bm{S}\rangle\cdot\bm{S})-\rho]+A_{sd}(\varphi-\rho)\nonumber\\
&+&R_{op}[\varphi(1+2\hat{\bm{s}}\cdot\bm{S})-\rho],
\end{eqnarray}
where $\varphi=\rho/4+\bm{S}\cdot\rho\bm{S}$ is the nuclear spin part which is unaffected by spin collisions~\cite{appelt98}, $\bm{B}=B_0\hat{\bm{y}}+B_{rf}\sin{\omega_{rf}t}\hat{\bm{z}}$, $A_{sd}$ is the spin-destructive collision rate, $\hat{\bm{s}}$ is the polarization of the optical pumping beam, and $\langle\bm{S}\rangle=\text{Tr}(\bm{S}\rho)$. Using the experimental parameters, we extract $\omega_L$ from the evolution of $\langle{S_x}\rangle$ in the simulation. To improve the  quantitative agreement between the simulation and  the experimental data, we have also tried to include an attenuation of 4\% of the rf field amplitude in the simulation without changing other independently calibrated parameters. This is probably due to the attenuation of rf fields by materials around the cell which makes us overestimate the rf amplitude using only the coil parameters calibrated in dc fields.  The results with and without the extra rf attenuation are shown in Fig.~\ref{fig:Ase}(c). We adopt the simulation results using this modified rf amplitude in the rest of this paper.

\begin{figure}[hbt]
\includegraphics[width=3in]{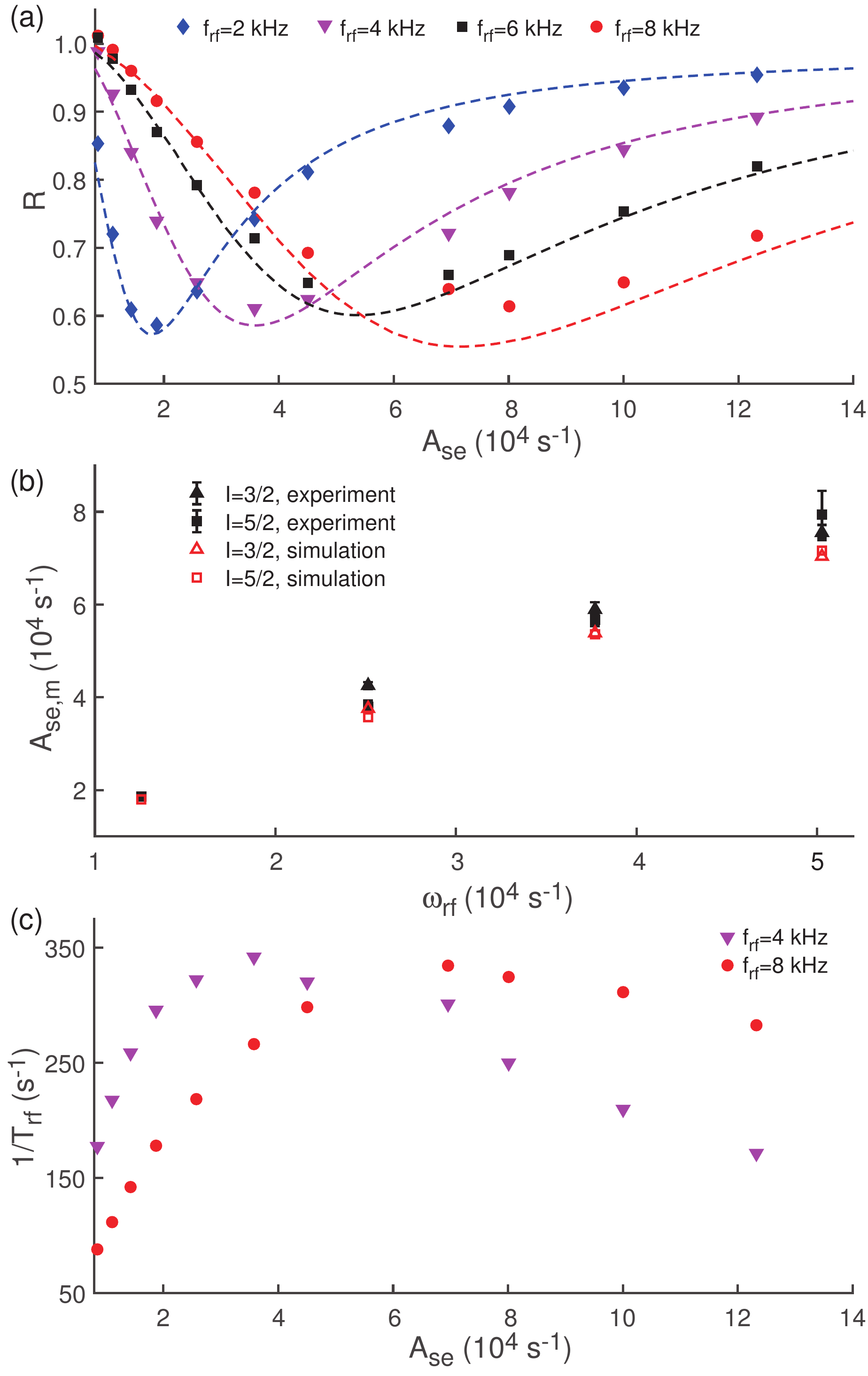}
\caption{\label{fig:freqsys}(Color online) (a) Experimental (solid points) and simulation results (dash lines) of $R$ as a function of $A_{se}$ for different rf field frequencies. The data is taken for atoms with nuclear spin $I=5/2$ at $B_0=0.3$~mG. (b) Experimental (solid points) and simulation results (open points) of $A_{se,m}$ at different rf field frequencies $\omega_{rf}$ for atoms with nuclear spin $I=3/2$ (triangle points) and $I=5/2$ (square points). (c) Experimental data of the depolarization rate due to the rf fields for $\omega_{rf}=2\pi\times 4$~kHz (nabla points) and 8~kHz (diamond points), and other parameters are the same as in plot~(a).}
\end{figure}

Several experimental parameters contribute to the modified rf spectroscopy in the strong spin-exchange collision regime. Fig.~\ref{fig:freqsys}(a) shows results at several different rf field frequencies. For clear comparisons, we have chosen $B_{rf}$ so that the minimum $R$ is around 0.6 for different $\omega_{rf}$. The results show that the curve shifts with $\omega_{rf}$. To better characterize this shift, we define $A_{se,m}$ as the value of $A_{se}$ at minimum $R$, and plot $A_{se,m}$ as a function of $\omega_{rf}$ in Fig.~\ref{fig:freqsys}(b). The data shows a linear relation between $A_{se,m}$ and $\omega_{rf}$ with a slope of 1.52, which agrees with the simulation result $A_{se,m}=1.42\omega_{rf}$ within 10\%.  

We further check the same behavior for atoms with nuclear spin $I=3/2$, which requires larger $A_{se}$ to reach the strong spin-exchange regime due to the larger atomic gyromagnetic ratio. Fig.~\ref{fig:freqsys}(b) shows the results of $A_{se,m}>3\times10^4$~s$^{-1}$ for atoms with $I=3/2$, and both the experimental data and the simulation results show a linear slope of 1.30. This value agrees with the $I=5/2$ system within 15\%. Compared with the factor of 2 difference in the gyromagnetic ratio in the SERF regime and a 50\% difference in nuclear spin slowing down factor of the transversal spin relaxation for such two systems, we conclude that the dependence of $A_{se,m}$ on the nuclear spin is weak.

We gain physics insight on this observation by  using a damped oscillator model. Without the rf fields, atoms follow a spin-temperature distribution due to the dominant spin-exchange collisions~\cite{happer72}. The rf field drives atomic transitions between Zeeman levels which disturb the spin-temperature distribution, and the spin-exchange collisions try to drive atoms back to the original distribution, entering as a damping factor for the oscillations. This damping effect increases linearly with $A_{se}$ when $A_{se}{\ll}\omega_{rf}$, and decreases inversely with $A_{se}$ when $A_{se}\gg\omega_{rf}$~\cite{shah09}. Therefore, we expect the damping effect to be strongest when $A_{se}$ is comparable with $\omega_{rf}$. Because the Larmor precession and rf transitions are coupled together in the  spin precessions, the damping effect described above affects the spin precessions in the same way. To characterize the damping effects due to the rf fields, we define the depolarization rate $1/T_{rf}$ by comparing the changes of $T$ in Eq.~\ref{eq:fit} with and without rf fields. Fig.~\ref{fig:freqsys}(c) shows the results for $1/T_{rf}$, which confirms the conclusion above. Due to the similarity with the damped oscillator model, we adopt such a model with the damping coefficient equal to $1/T_{rf}$, and the damping ratio is proportional to the ratio of $1/T_{rf}$ to the oscillation frequency.  For the oscillating components in Eq.~\ref{eq:fit}, we only need to consider the $\omega_L$ term, because the frequencies in other terms are much larger than $\omega_L$ and damping ratios are negligible. In the underdamped oscillator model, the oscillation frequency decreases as the damping ratio increases. Minimum values of $\omega_L$ and $R$ appear in the region with the largest damping rate, where $A_{se}$ is comparable with $\omega_{rf}$.

\begin{figure}[htb]
\includegraphics[width=3in]{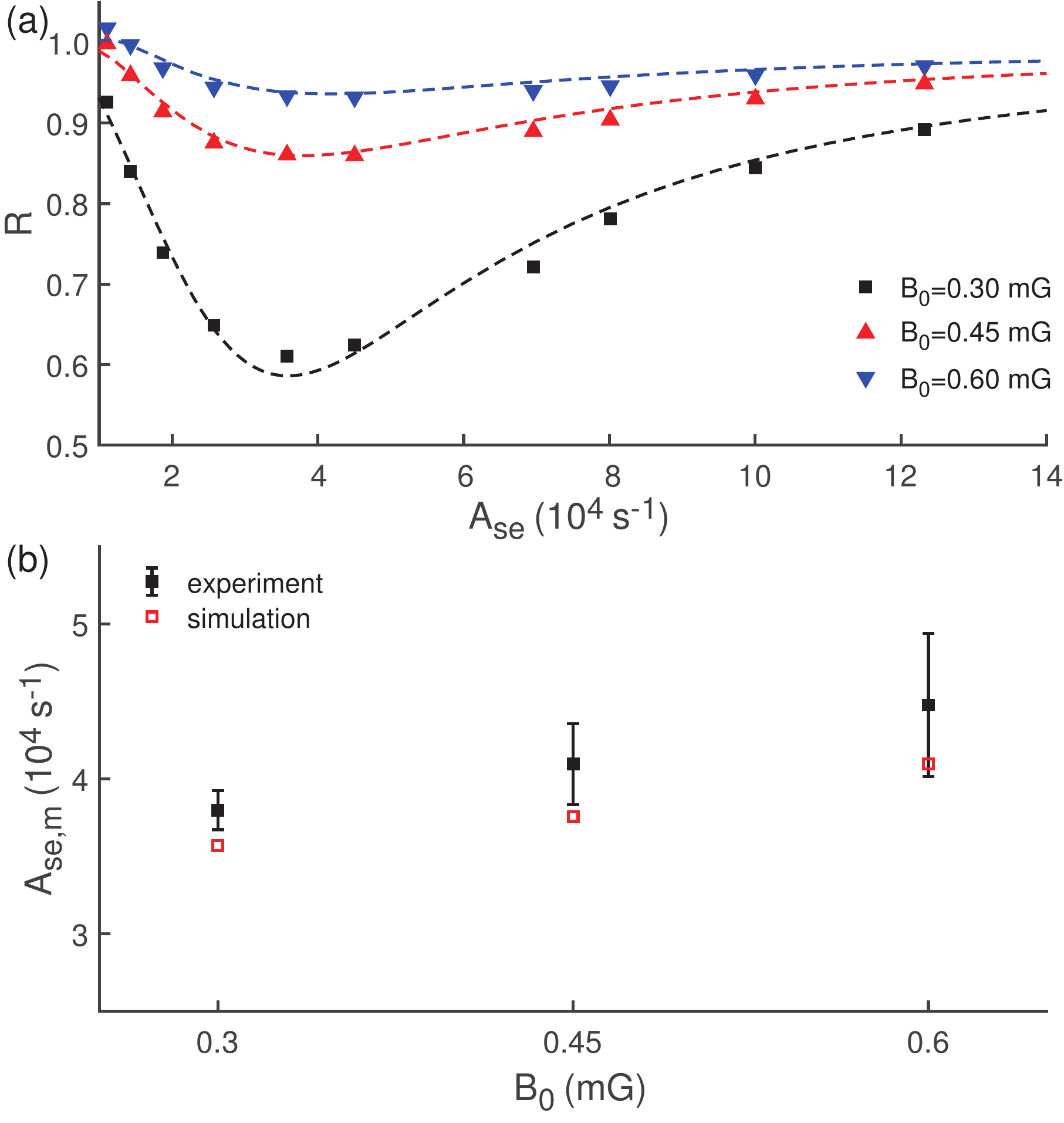}
\caption{\label{fig:biasmag}(Color online) (a) Experimental (solid points) and simulation (dash lines) results of $R$ as a function of $A_{se}$  at different bias fields $B_0$, where $I=5/2,$ $\omega_{rf}=2\pi\times4$~kHz, and $\gamma_0B_{rf}/\omega_{rf}=0.391$. (b) Experimental (solid points) and simulation (empty boxes) of $A_{se,m}$ for the plots in plot~(a).}
\end{figure}

According to the model above, an increase in Larmor frequency $\omega_L$ reduces the damping ratio, and results in a decreased modification on $R$. Fig.~\ref{fig:biasmag}(a) shows the experimental and simulation results for $R$ as a function of $A_{se}$ at different $B_0$, with other parameters the same as those in Fig.~\ref{fig:Ase}(b). We find that the contrast of the curve decreases by a factor of 6 as $B_0$ increases from 0.3~mG to 0.6~mG, such a rapid decrease in the contrast would lead to no experimentally observable relation between $R$ and $A_{se}$ when $B_0$ is larger than 1~mG. However, $A_{se,m}$ is weakly dependent on $B_0$ as shown in Fig.~\ref{fig:biasmag}(b), where $A_{se,m}$ changes less than 20\% when  $B_0$ increases from 0.3~mG to 0.6~mG.

\section{Large rf field amplitude regime}

\begin{figure}[htb]
\includegraphics[width=3in]{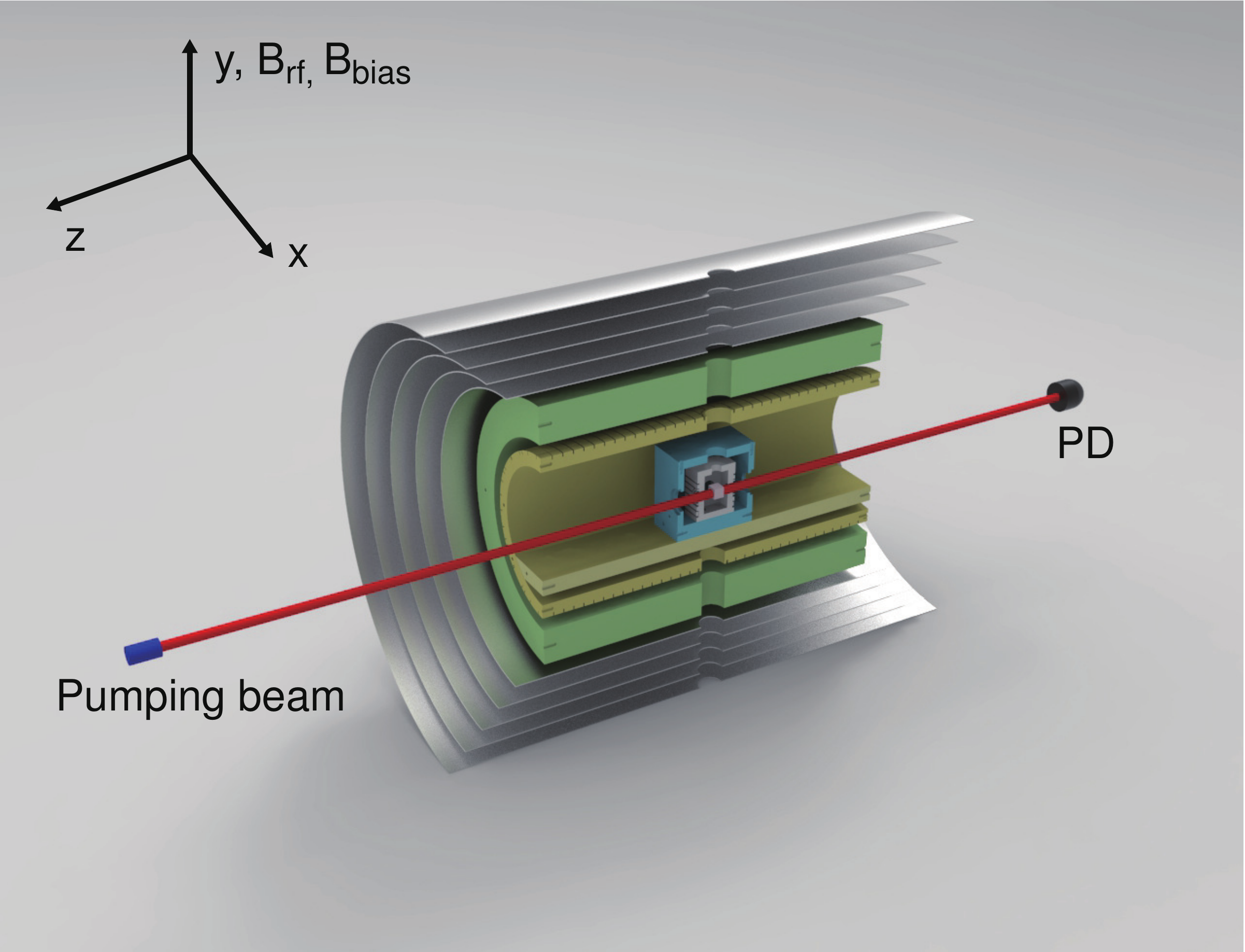}
\caption{\label{fig:setup2}(Color online) Zero-field level-crossing resonance scheme for measurements at large rf field amplitude.}
\end{figure}

When the rf field amplitude is large, especially when the modulation index parameter $\gamma_0B_{rf}/\omega_{rf}>1$, the experiment and data analysis methods described in the previous section are invalid. To study the effects in this regime, we use the zero-field level-crossing resonance scheme~\cite{cohen70,mhaskar12,sheng17}. As shown in Fig.~\ref{fig:setup2}, we continuously apply rf fields along the $y$ direction, and monitor $S_z$, the electronic spin polarization in the $z$ direction, through the pumping beam absorption. 

\begin{figure}[htb]
\includegraphics[width=3in]{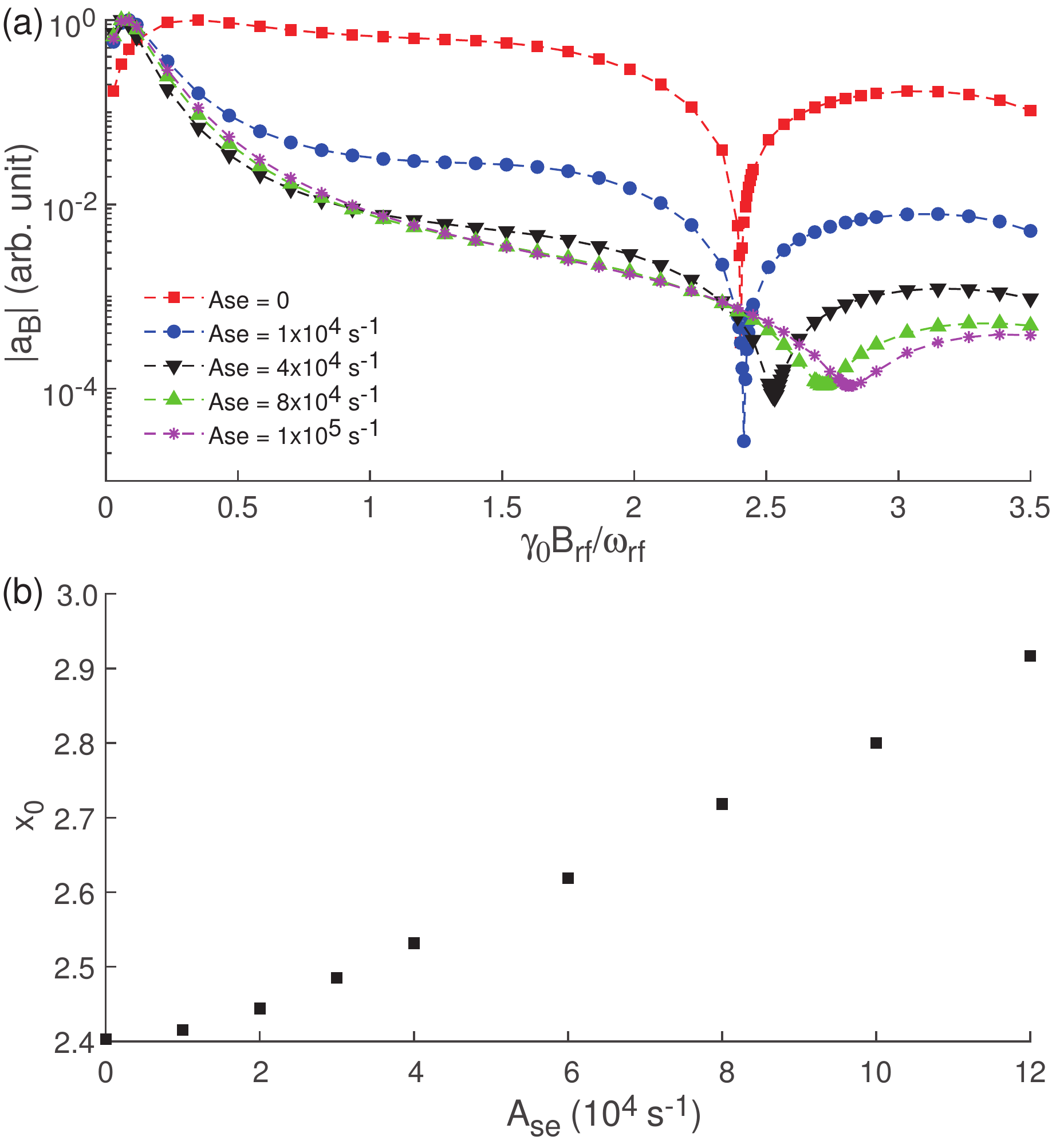}
\caption{\label{fig:zero}(Color online) (a) The absolute values of simulated $a_B$ as a function of rf field parameters, with $B_y=0.01$~mG and $\omega_{rf}=2\pi\times4$~kHz. (b) The value of $x_0$ as a function of $A_{se}$.}
\end{figure}

We simulate the system using the same density matrix methods as in Sec. II, where the correlation between different hyperfine states are set to zero after the actions of spin-exchange and other terms. When the amplitude of the bias field $B_y$ is small, the modulation component of $S_z$ at $\omega_{rf}$ is proportional to $B_y$~\cite{shah09}, that is, $S_z(w=\omega_{rf})=a_BB_y$, where $a_B$ is proportional to atomic gyromagnetic ratio and depolarization time.  Fig.~\ref{fig:zero}(a) shows the simulation results of $|a_B|$, where we focus on the dip in the curve of $|a_B|$. For the small region around the dip, we could treat the depolarization time as a constant. Then such a dip corresponds to a minimum gyromagnetic ratio point due to the rf photon dressing. When $A_{se}=0$, this point corresponds to the first zero point of $J_0(\gamma_0B{rf}/\omega_{rf})$ in Eq.~1. As explained in Sec. II, spin-exchange collisions work as counter effects on atomic transitions compared with rf fields. Denoting the value of $\gamma_0B_{rf}/\omega_{rf}$ at minimum $|a_B|$ by $x_0$, we should expect $x_0$ to increase with $A_{se}$, which agrees with the simulation results as shown in Fig.~\ref{fig:zero}(b). 

\section{conclusion}
In conclusion, we have studied the interactions between non-resonant oscillating rf fields and atoms with strong spin-exchange collisions. In the weak rf field amplitude regime, we observe a strong modification of the atomic Larmor frequency when the collision rate and the rf field frequency are comparable ($A_{se}\sim1.4\omega_{rf}$), which is also confirmed by density matrix simulations. The amplitude of this modification decreases rapidly when the bias field increases. This effect has been unnoticed before probably due to its narrow observation window. This new phenomena is  explained by an underdamped oscillator model, and the damping effect comes from the disturbance of the spin-temperature distribution determined by the strong spin-exchange collisions. Atomic magnetometers based on the same system studied in this paper have emerged to be promising and complementary tools in biomagnetic applications~\cite{boto18}. Identification of these effects is helpful in calibrations of atomic gyromagnetic ratio and spin-exchange collision rates in such systems. In the large rf field amplitude regime, strong spin-exchange interactions change the position of the minimum gyromagnetic ratio point, where atoms are least sensitive to magnetic fields. A system working on such a point could be used for spin control related applications.

\section{Acknowledgements}
We thank L. A. Orozco for encouragement and comments on the manuscript, and Z.-T. Lu for discussions. This work was supported by Natural Science Foundation of China (Grant No.11774329) and Thousand Youth Talent Program of China.

\end{document}